\documentstyle[twocolumn,prl,aps,epsf]{revtex}
\begin{document} %\draft
\title{
Electron Localization in a 2D System with Random Magnetic Flux
}
\author{D. Z. Liu$^{1,}$\cite{addr}, X. C. Xie$^{1,}$\cite{axie}, S. Das
%SCZ
Sarma$^1$ and S. C. Zhang$^2$}
\address{$^1$
Center for Superconductivity Research,
Department of Physics, University of Maryland,
College Park, MD 20742
}
\address{$^2$
Department of Physics, Stanford University, Palo Alto, CA
94305}
\address{\rm (Submitted to Physical Review Letters on 26 August 1994)}
\address{\mbox{ }}
%\address{\mbox{ }}
%\begin{abstract}
\address{\parbox{14cm}{\rm \mbox{ }\mbox{ }
Using a finite-size scaling method, we calculate the localization
properties of a disordered two-dimensional electron system in the
presence of a random magnetic field.
%SCZ
%A mobility edge is observed that
Below
a critical energy $E_c$
all states are localized and the localization length $\xi$ diverges
when the Fermi energy approaches the critical energy, {\it i.e.}
$\xi(E)\propto |E-E_c|^{-\nu}$. We find that $E_c$
shifts with the strength of the disorder and the amplitude of the
random magnetic field while the critical exponent ($\nu\approx 4.8$)
remains unchanged indicating universality in this system.
Implications
on the experiment in half-filling fractional quantum Hall system are also
discussed.
}}
\address{\mbox{ }}
%\address{\mbox{ }}
%\end{abstract}
\address{\parbox{14cm}{\rm PACS numbers: 72.10.Bg, 71.55.Jv, 72.15.Rn,
73.40.Hm}}
\maketitle

\makeatletter
\global\@specialpagefalse
\def\@oddhead{REV\TeX{} 3.0\hfill Das Sarma Group Preprint, 1994}
\let\@evenhead\@oddhead
\makeatother

%%%%%%%%%%%%%%%% Introduction %%%%%%%%%%%%%%%%%%%%%%%%%%%%
Fractional quantum Hall (FQH) system \cite{prange:qhe} is an ideal
candidate to study
localization properties in strongly correlated electron systems.
In a non-interacting picture,
according to the scaling theory of localization \cite{anderson:scal}
all electrons in a two-dimensional system are localized in the absence of
magnetic field. When the two-dimensional electron system is subject to
a strong perpendicular magnetic field, the energy spectrum becomes
a series of impurity broadened Landau levels.
%SCZ: there is a finite density of extended states (not only one)
Extended state appear in the center of each Landau band,
%SCZ: not so simple: (due to the breaking time-reversal invariance)
while states at other energies are localized.
This gives rise to the integer
quantum Hall effect. It has been demonstrated that the critical
transitions in the center of each Landau band are
universal\cite{ldz:bloc}.
In the FQH regime, electron-electron interaction plays an
important role \cite{prange:qhe}, however, recent experiments and theories
%SCZ: thus the localization properties are quite different
%from the non-interacting picture.
indicate that the critical properties of the plateau transition might also
be in the same universality class\cite{tsui,klz}.

Recently, Halperin, Lee and Read \cite{halperin}
and Kalmeyer and Zhang\cite{scz-1} developed an effective Chern-Simons
field theory
to understand electronic properties of the FQH systems.
In their theory the quasi-particles are weakly interacting
composite fermions (as originally proposed by Jain \cite{jain:com})
which can be
constructed by attaching an even number of flux quanta to electrons under a
Chern-Simons transformation. In this simple picture, the fractional quantum
Hall effect can be mapped into the {\it integer} quantum Hall effect for
the composite fermion system subject to an effective magnetic
field \cite{jain:com}.
%SCZ
%It then follows that localization properties at
%magic
%fractional filling factors can be mapped to those of integer Landau
%levels \cite{scz-2}.
At filling factor $\nu_f=\frac{1}{2}$,
although the effective magnetic field  $B^*$ vanishes,
composite fermions are subject to the
%SCZ
random fluctuations of the gauge field, induced by the ordinary impurities
\cite{halperin,scz-1}.
Thus, it is important to
study the localization properties of non-interacting charged particles
in a random magnetic field to understand the half-filling FQH system.
The problem of charged particles moving in a random magnetic field
is also relevant to the theoretical studies of
high $T_c$ models
%SCZ
where the gauge filed fluctuations play an important role
\cite{lee-1}.

%SCZ
According to the conventional scaling theory of localization, the
random flux system belongs to the unitary ensemble, which is described
by a nonlinear sigma model with unitary symmetry. Since there is
no net magnetic field, the topological term of the uniform
magnetic field case is absent.
Perturbative
renormalization group calculations show that
all states are localized \cite{hikami}. However, it has been argued recently by
Zhang and Arovas\cite{za} that although the constant topological term
is absent, there is a term describing the long ranged interaction between
the topological densities, and they conjectured that this new term
could lead to a phase transition from localized to extended states.
There have been a number of conflicting numerical investigations on the
localization properties of the random magnetic field system. The
conclusions in these studies range from all states
localized\cite{jap-1}
%SCZ: we never claimed that all states are localized, it was mis-quoted
%in Nagaosa's paper.
to extended states around band center\cite{scz-1,jap-2}.
In Ref.\cite{scz-2}, the authors
found evidence for a
mobility edge, but their system is neither large enough to see  good
scaling nor close enough to the critical regime to obtain a conclusive
critical exponent.

In this paper, we systematically investigate the localization
properties of a disordered two-dimensional electron system in the
presence of a random magnetic field. The localization length is
calculated using a transfer matrix technique and finite-size scaling
analysis.
An important strength of our calculation is using system widths (upto
128) which are substantially (by a factor of four) larger than
those\cite{jap-1,jap-2,scz-2} existing in the literature.
We find the following results:
(1) A mobility edge $E_c$ is observed
and the localization length $\xi$ diverges
when the Fermi energy approaches the critical energy;
(2) we find that the critical energy $E_c$
shifts with increasing  disorder strength;
(3) $E_c$ shifts with changing
randomness in the magnetic field;
(4) the critical exponent ($\nu\approx 4.5$) remains unchanged while
varying the disorder strength and the randomness of the magnetic field,
indicating universality in the metal-insulator transition;
(5) the mobility edge survives in the presence of weak but non-zero
average random magnetic field.

%%%%%%%%%%%%%%% Model and Method %%%%%%%%%%%%%%%%%%%%%%%%%
We model our two-dimensional system in a very long strip geometry
with a finite width ($M$) square
lattice
with nearest neighbor hopping. The disorder potential is modeled by
the on-site white-noise potential $V_{im}$ ($i$ denotes the column
index, $m$ denotes the chain index) ranging from $-W/2$ to $W/2$.
Random magnetic field is introduced by varying the flux in each lattice
plaquette uniformly between $-\phi_r/2$ and $\phi_r/2$ (in this case
the average field is zero, we also discuss the situation of weak but non-zero
average random magnetic field in the later part of this paper). The
Hamiltonian of this system can be written as:
\begin{eqnarray}
{\cal H} &= &\sum_i\sum_{m=1}^{M} V_{im}|im><im| \\
& &+\sum_{<im;jn>}\left[
t_{im;jn}|im><jn|+t^{\dagger}_{im;jn}|jn><im|\right],\nonumber
\end{eqnarray}
where $<im;jn>$ indicates nearest neighbors on the lattice. The
amplitude of the hopping term is chosen as the unit of the energy.
A specific gauge is chosen so that the inter-column
hopping does not carry a complex phase factor ({\it i.e.}
$t_{im;i+1,m}=-1$).  The only effect of
random magnetic field shows up on the phase factor of the intra-column
(inter-chain)
hopping term. If the random flux in a plaquette cornered by
$(im), (i+1,m), (i+1,m+1)$ and $(i,m+1)$ is $\phi_{im}$, then
\begin{equation}
\frac{t_{i+1,m;i+1,m+1}}{t_{im;i,m+1}}=\exp \left[i2\pi
\frac{\phi_{im}}{\phi_o}\right],
\end{equation}
where $\phi_o=hc/e$ is the magnetic flux quantum.
For a specific energy $E$, a transfer matrix $T_{i}$ can be easily set up
mapping the wavefunction amplitudes at column ${i-1}$ and $i$ to those
at column $i+1$, {\it i.e.}
\begin{equation}
\left( \begin{array}{c} \psi_{i+1} \\ \psi_{i} \end{array} \right) =
T_{i}
\left( \begin{array}{c} \psi_{i} \\ \psi_{i-1} \end{array} \right) =
\left( \begin{array}{cc} H_{i}-E & -I \\ I & 0 \end{array} \right)
\left( \begin{array}{c} \psi_{i} \\ \psi_{i-1} \end{array} \right) ,
\end{equation}
where $H_i$ is the Hamiltonian for the $i$th column, $I$ is a $M\times
M$ unit matrix.
Using a standard iteration algorithm \cite{ldz:bloc}, we can
calculate the Lyapunov exponents for the transfer matrix $T_{i}$.
 The localization length $\lambda_M(E)$ for energy
$E$ at finite width $M$ is then given by the inverse of the smallest
Lyapunov
exponent. In our numerical calculation, we choose the sample length to
be over $10^4$ so that the self-averaging effect automatically takes
care of the ensemble statistical fluctuations. A sample of our
calculated finite width localization length for various energies is
shown in Fig.\ref{nuhf-f1}(a).

We use the standard one-parameter finite-size scaling analysis
\cite{ldz:bloc} to
obtain the thermodynamic localization length $\xi$. According to the
one-parameter scaling theory, the renormalized finite-size
localization length $\lambda_M/M$ can be expressed in terms of a
universal function of $M/\xi$,
{\it i.e.}
\begin{equation}
\frac{\lambda_M(E)}{M}=f\left( \frac{M}{\xi(E)}\right),
\end{equation}
where $f(x) \propto 1/x$ in thermodynamic limit $M\rightarrow \infty$
while approaching a constant($\sim 1$) at the mobility edge where the
thermodynamic localization length diverges. Numerically we shift the
data in Fig.\ref{nuhf-f1}(a) onto a smooth function

%%%%%%%%%%%%%%%%%%%%%%%%%%% Fig.1 %%%%%%%%%%%%%%%%%%%%%%%%%%%%%%%%%%%
\begin{figure}
%Fig.1a
%\epsfbox{/nfs/rainbow/o/ldz/tek/paper/phd11/f1a.ps}
%\epsfig{file=f1a.ps,height=6.0cm}
 \vbox to 6.0cm {\vss\hbox to 8cm
 {\hss\
   {\includegraphics{/nfs/rainbow/o/ldz/tek/paper/phd11/f1a.ps}
   }
  \hss}
 }
\vspace{3mm}
%Fig.1b
 \vbox to 6.0cm {\vss\hbox to 8cm
 {\hss\
%\epsfbox{/nfs/rainbow/o/ldz/tek/paper/phd11/f1b.ps}
   {\includegraphics{/nfs/rainbow/o/ldz/tek/paper/phd11/f1b.ps}
   }
  \hss}
 }
\caption{(a) Renormalized finite-size localization length ($\lambda_M/M$)
for different sample width $M$ in a random magnetic
field($\phi_r=1.0$) without on-site disorder ($W=0$); (b) Scaling
function and localization length in the thermodynamic limit(inset)
with $\nu =4.52\pm 0.08$ and $E_c=-3.00$. Here different symbols
represent different energies.
\label{nuhf-f1}}
\end{figure}

\noindent
with a least
square fit. Note that we have to select data for a large enough sample
in the scaling analysis so as to avoid severe finite-size
effect. In our calculation, we choose the data for sample width
greater than 16 (inclusive).
The thermodynamic localization length is given by the
amount of shifts on a log-log plot. A sample of the scaling function
and corresponding thermodynamic localization length is shown in
Fig.\ref{nuhf-f1}(b). Because of the symmetry in the problem, we only
study the branch with negative energy $E<0$.

%%%%%%%%%%%%%%% Numerical Result %%%%%%%%%%%%%%%%%%%%%%%%%
We first study the case with random magnetic field characterized by
fluctuation amplitude $\phi_r=1.0$.
We find that if the Fermi energy is below $E_c=-3.0$, the finite-size
localization length is well converged and always smaller than the
sample width indicating that all states are localized below $-3.0$. On
the contrary, for the electronic states with energy
higher than $-3.0$ the inverse of the Lyapunov exponent is always
larger than the sample width which is the feature of extended states.

%SCZ: following statements are not true. Even though time reversal is broken,
%and leading order weak localization effects are destroyed, the higher order
%correction could still localize all states. Thus T-breaking alone does not
%lead to extended states. One needs a new term in the nonlinear sigma model
%e.g. a long ranged interaction between the topological densities.
%Presence of a random magnetic field has a two-fold effect in a
%two-dimensional system. (i) It introduces randomness in the
%system (even without a non-magnetic disorder potential), thus
%electronic states at the bottom of the band with smaller scattering
%space become localized; (ii) It
%breaks the local time-reversal symmetry even though the overall average
%magnetic field is zero, hence the weak localization mechanism is
%destroyed resulting in extended states at higher energy. A mobility edge
%is a natural consequence of applying a random magnetic field.

%%%%%%%%%%%%%%%%%%% Fig.2 %%%%%%%%%%%%%%%%%%%%%%%%%%%%%%%%%%%%%%
\begin{figure}
%Fig.2a
%\epsfbox{/nfs/rainbow/o/ldz/tek/paper/phd11/f2a.ps}
 \vbox to 6.0cm {\vss\hbox to 8cm
 {\hss\
   {\includegraphics{/nfs/rainbow/o/ldz/tek/paper/phd11/f2a.ps}
   }
  \hss}
 }
\vspace{3mm}
%Fig.2b
%\epsfbox{/nfs/rainbow/o/ldz/tek/paper/phd11/f2b.ps}
 \vbox to 6.0cm {\vss\hbox to 8cm
 {\hss\
   {\includegraphics{/nfs/rainbow/o/ldz/tek/paper/phd11/f2b.ps}
   }
  \hss}
 }
\caption{Scaling functions and thermodynamic localization length
(inset) in
in random magnetic fields with (b) or without(a) on-site disorder. The
corresponding critical exponents and critical energies are: (a) $\nu
=4.98\pm 0.10, E_c=-3.13$; (b) $\nu =4.86\pm 0.02, E_c=-2.73$.
\label{nuhf-f2}}
\end{figure}

\noindent
We also find
that the thermodynamic localization length $\xi$ diverges while approaching
$E_c=-3.0$ indicating the existence of
a mobility edge around $-3.0$.
Our best fit analysis indeed gives a critical energy $E_c=-3.00$ with
a critical exponent $\nu=4.52$ appearing in $\xi =|E-E_c|^{-\nu}$ for
$E < E_c$ (the critical exponent is given by the slope of the
straight line in the inset log-log plot of Fig.\ref{nuhf-f1}(b)).

We expect that the mobility edge should shift to lower energy if the
magnetic field is less random so that extended states are more
favorable. This is exactly what we observe in our calculation.
In Fig.\ref{nuhf-f2}(a) we show the
typical scaling function and localization length for a less random
magnetic field with $\phi_r=0.9$. In Table \ref{nuhf-t1}, we present
the critical exponents and
critical energies for different randomness in the
magnetic field. We find that the critical energy increases almost
linearly with the random flux amplitude which is proportional to the
energy fluctuation created by the random magnetic field.

We now discuss the effect due to correlation between random magnetic
field and disorder potential.
We consider two types of disorder
potential: (i) Independent model where the distribution of disorder
potential is completely independent of the random magnetic field. This
model is relevant for the case with random distributed non-magnetic
impurities in the sample. (ii) Correlated model in which the strength
of disorder is associated with the local random magnetic field
(numerically we select each on-site disorder so that it is proportional
to the random flux in the neighboring plaquette). This
model is relevant for the case with random distributed magnetic
impurities, for example, disorder-pinned random flux lines in the
sample.

In the independent model, since non-magnetic random disorder potential
tends to localize all the
electronic states, we expect that the mobility edge in a random
magnetic field should  shift to higher energy with increasing strength
of disorder potential because the localized states are more favorable
in this situation.
In Fig.\ref{nuhf-f2}(b) we show the
typical scaling function and localization length
 for a strongly
disordered sample with $W=2.0$ in a random magnetic field with $\phi_r=1.0$.
As presented in Table \ref{nuhf-t1}, the critical energy moves to
higher energy as disorder strength $W$ increases for fixed random
magnetic field ($\phi_r=1.0$) just as we expect.
Presented in the last row of Table \ref{nuhf-t1} is the data for a
correlated disorder model, the behavior of the mobility edge is quite
similar to that for the independent disorder model.

It is interesting to notice that even though the mobility edge is
shifting for various random magnetic field and
disorder strength and correlation, the critical exponent
for  the metal-insulator transition
is more or less unchanged which indicates the
universality of this critical transition. Our calculated exponent
($\nu\approx 4.5$) is quite different

%%%%%%%%%%%%%%%%%%% Table 1 %%%%%%%%%%%%%%%%%%%%%%%%%%%%%%%%%%%%%%%%%
\begin{table}
\caption{The critical exponent($\nu$) and critical energy($E_c$) for
different representations of random magnetic field ($\phi_r$) and on-site
disorder ($W$). The last row is for correlated disorder model while
the other data are for the independent disorder model.
\label{nuhf-t1}}
\begin{tabular}{cccc}
 $\phi_r$ & $W$  & $\nu$ & $E_c$ \\\hline
0.7 & 0.0 & $4.53\pm 0.11$ & -3.40 \\
0.8 & 0.0 & $4.60\pm 0.06$ & -3.28 \\
0.9 & 0.0 & $4.98\pm 0.10$ & -3.13 \\
1.0 & 0.0 & $4.52\pm 0.08$ & -3.00 \\
1.0 & 1.0 & $4.79\pm 0.17$ & -2.85 \\
1.0 & 2.0 & $4.86\pm 0.02$ & -2.73 \\
1.0 & 3.0 & $5.29\pm 0.60$ & -2.13 \\
0.8 & 2.0 & $4.45\pm 0.10$ & -3.35
\end{tabular}
\end{table}

\noindent
from the value obtained in
Ref.\cite{scz-2}. However, our system is much larger (4 times larger)
than the system
studied in Ref.\cite{scz-2}, our scaling is much better and our data
are closer to the critical regime, thus our calculated exponent better
represents the true critical exponent.

%%%%%%%%%%%%%%%%%%%%% Fig.3 %%%%%%%%%%%%%%%%%%%%%%%%%%%%%%%%%%%%%%%
\begin{figure}
%Fig.3
%\epsfxsize=0.3
%\epsfbox{/nfs/rainbow/o/ldz/tek/paper/phd11/f3.ps}
 \vbox to 6.0cm {\vss\hbox to 8cm
 {\hss\
   {\includegraphics{/nfs/rainbow/o/ldz/tek/paper/phd11/f3.ps}
   }
  \hss}
 }
\caption{
Thermodynamic localization length in the presence of a weak but non-zero
average random magnetic field.
\label{nuhf-f3}}
\end{figure}

In Fig.\ref{nuhf-f3} we present the localization length for a
two-dimensional system subject to a random magnetic field when
the average field is weak but non-zero ({\it i.e.} $<\phi>\neq
0$). Clearly there exists a mobility edge at $Ec\approx -3.0$ for
$<\phi>=0.01$.
It is consistent with the argument that in the vicinity
of $<B^*>=0$, the composite fermion system behaves as a
Fermi liquid.
%SCZ: Jain's argument is based on clean systems, I do not believe there
%is any general argument for the dirty case. Please check!

%%%%%%%%%% SDS Insertion %%%%%%%%%%%%%%%
As mentioned earlier there is currently considerable disagreement in
the literature about the nature of two-dimensional one-electron
eigenstates in a random flux environment. An earlier finite size
scaling analysis \cite{jap-1} concluded that all states are localized,
but the localization lengths are exponentially large near the
band center. Our largest system widths (=128) are four times larger
than those (=32) used in Ref.\cite{jap-2} and our results are
consistent with the existence of mobility edges separating localized
and delocalized states with a localization exponent $\nu\approx 4.5$.
We emphasize, however, that no finite-size scaling analysis can
distinguish between extended states and states with extremely large
(orders of magnitude larger than the system length in Ref.\cite{jap-1})
localization lengths \cite{jap-1,aronov}. We contend that experiments
in real samples cannot distinguish between these two scenarios either
{\it i.e.} the situation with extremely large localization lengths
would behave very much like a weakly dirty metal. The important point
to realize is that the finite-size scaling evidence in favor of a
delocalization transition presented in this paper is as strong as it
is in the corresponding quantum Hall plateau
transitions \cite{ldz:bloc} where the existence of extended
states is not in doubt. We have also calculated the conductance $g$ as
a function of the system length for various (Fermi) energies using a
direct transfer matrix Landauer formula type approach, finding results
completely consistent with those presented here ({\it i.e.} an
insulator for $E<E_c$ and a metal for $E>E_c$). Our efforts to
construct a beta-function $\beta(g)$ have not, however, yielded
unambiguous results because of inherent fluctuations in the calculated
conductance. Systems much larger than those used here ($M>>128$),
which are not accessible with the currently available computers, will be
needed to obtain an unambiguous $\beta(g)$ -- in that sense, we have a
disagreement here with the conclusion of Ref.\cite{jap-1} and
\cite{aronov}.

%%%%%%%%%%%%%%% Conclusion %%%%%%%%%%%%%%%%%%%%%%%%%%%%%%%

%This universal critical phenomena could be verified by the FQHE experiment
%at the vicinity of half-filling where the model of weakly interactive
%%composite
%fermions subject
%to randomly fluctuating gauge field is relevant. Since the mobility edge
%shifts to higher energy by increasing the strength of the disorder
%potential, transport measurements on different samples with different
%impurity doping level should give qualitatively different results
%depending on the location of the Fermi level, namely in the extended
%regime or the localized regime. If the Fermi level is in the extended
%regime, the half-filling FQH system should have metallic behavior;
%activated behavior should be observed if the Fermi level enters the
%localized regime.
%SCZ: this has already been done in H.W. Jiang's group at UCLA
%We propose a measurements of transport properties of
%half-filling FQH system under a backgate which could adjust the
%impurity potential strength in the same sample.

We thank J. K. Jain, T. Kawamura, and T. R. Kirkpatrick for helpful
discussions.
This work is supported by the U.S. National Science Foundation.

%%%%%%%%%%%%%%% Reference %%%%%%%%%%%%%%%%%%%%%%%%%%%%%%%%


\begin{references}

\bibitem[*]{addr} Current address: The James Franck Institute, The
University of Chicago, Chicago, IL 60637.

\bibitem[**]{axie} Present and permanent address:
Department of Physics, Oklahoma State University,
Stillwater, OK 74078.

\bibitem{prange:qhe} For a review, see {\it The Quantum Hall Effect},
edited by R.E. Prange and S.M. Girvin (Springer-Verlag, New York, 1990).

\bibitem{anderson:scal} E. Abrahams, P.W. Anderson, D.C. Licciardello,
and T.V. Ramakrishnan,
Phys. Rev. Lett. {\bf 42}, 673 (1979); D. Belitz and T.R. Kirkpatrick,
Rev. Mod. Phys. {\bf 66}, 261 (1994).

\bibitem{ldz:bloc} D.Z. Liu and S. Das Sarma, Phys. Rev. B, {\bf 49},
2677 (1994), and references therein.

%SCZ
\bibitem{tsui} H.P. Wei et al, Phys. Rev. Lett. {\bf 61}, 1294 (1988).

%SCZ
\bibitem{klz} S. Kivelson and D.H. Lee and S.C. Zhang, Phys. Rev. B
{\bf 46}, 2223 (1992).

\bibitem{halperin} B.I. Halperin, P.A. Lee and N. Read, Phys. Rev. B
{\bf 47}, 7312 (1993).

\bibitem{scz-1} V. Kalmeyer and S.C. Zhang, Phys. Rev. B {\bf 46},
9889 (1992).

\bibitem{jain:com} J.K. Jain, Phys. Rev. Lett. {\bf 63}, 199 (1989).

\bibitem{lee-1} N. Nagaosa and P.A. Lee, Phys. Rev. Lett. {\bf 64},
2450 (1990), and references therein.

\bibitem{hikami} S. Hikami, Prog. Theor. Phys. Suppl. {\bf 107}, 213
(1992).

\bibitem{za} S. C. Zhang and D. Arovas, Phys. Rev. Lett. {\bf 72},
1886 (1994).

\bibitem{jap-1} T. Sugiyama and N. Nagaosa, Phys. Rev. Lett. {\bf 70},
1980 (1993).

\bibitem{jap-2} Y. Avishai and Y. Hatsugai and M. Kohmoto,
Phys. Rev. B {\bf 47}, 9561 (1993).

\bibitem{scz-2} V. Kalmeyer, D. Wei, D.P. Arovas, and S.C. Zhang,
Phys. Rev. B {\bf 48}, 11095 (1993), and references therein.

\bibitem{aronov} A.G. Aronov, A.D. Mirlin, and P. W\"{o}lfle, Phys. Rev. B
{\bf 49}, 16609 (1994).


\end{references}
\end{document}